\documentclass [english,aps,prb,twocolumn,superscriptaddress,amsmath,amssymb,showpacs,showkeys]{revtex4}
\usepackage[T1]{fontenc}
\usepackage[latin1]{inputenc}
\usepackage{graphicx}

\makeatletter

\providecommand{\tabularnewline}{\\}

\usepackage{epsfig}

\usepackage{prettyref}

\usepackage{babel}
\makeatother

\begin{document}

\title{Unusual non-equilibrium behavior of cupric oxide nanoparticles}

\author{Vijay Kumar Bisht}

\email{vijayb@iitk.ac.in}

\affiliation{Department of Physics, Indian Institute of Technology Kanpur 208016,
India}

\author{K.P. Rajeev}

\email{kpraj@iitk.ac.in}

\affiliation{Department of Physics, Indian Institute of Technology Kanpur 208016,
India}

\author{Sangam Banerjee}

\affiliation{Surface Physics Division, Saha Institute of Nuclear Physics Kolkata, 700 064, India}

\begin{abstract}

We report studies on temperature, field and time dependence of
magnetization on cupric oxide nanoparticles of sizes 9~nm, 13~nm and 16~nm.  The nanoparticles show unusual features 
in comparison to other antiferromagnetic nanoparticle systems. The  field
cooled (FC) and  zero field cooled (ZFC)  magnetization curves
bifurcate well above the Neel temperature and  the usual peak
in the ZFC magnetization curve is absent. 
 The system does not show any memory effects which is in sharp contrast to the usual behavior shown by other antiferromagnetic nanoparticles.  It turns out that the nature of CuO nanoparticles is very strange and is neither superparamagnetic nor spin glass-like .
\end{abstract}



\maketitle

\section{INTRODUCTION}
Magnetism in nanoparticles have been investigated intensively in the
last few decades because of their technological importance as well
as for understanding the physics involved in their many unusual properties vis-a-vis the bulk.\cite{Batlle} In a nanoparticle the magnetic properties are strongly affected by the large proportion of surface spins which face an entirely different environment in comparison to the particle's core.  Generally these systems show non-equilibrium behavior at low temperature with features such as a bifurcation in field cooled (FC) and zero field cooled (ZFC) susceptibility, slow  relaxation of magnetization, aging and memory effects.\cite{Sasaki,Sun,Malay,Chakraverty,Tsoi,RZheng,Martinez}
It is widely believed that such non-equilibrium behavior exhibited by magnetic nanoparticles can arise
mainly due to three mechanisms.  First, in non interacting nanoparticle
systems one can have superparamagnetism which arises from anisotropy energy barrier of each nanoparticle moment.\cite{Batlle,Sasaki,Neel,Brown,Malay,Tsoi} Second, in interacting nanoparticle systems, one can have superspin glass behavior which arises from   the frustration caused by competing dipolar interactions of neighboring particles coupled with the  randomness in particle positions and orientations of anisotropy
axes.\cite{Batlle,Sasaki,Sun,Malay} A third mechanism for non-equilibrium behavior has been proposed based on spin glass behavior arising due to the freezing of surface spins in a  nanoparticle caused by disorder at its surface. \cite{Martinez,Kodama,Tiwari,Winkler}

 Transition metal monoxides such as NiO, MnO, CoO, CuO etc. are all antiferromagnetic and nanoparticles
 of most of them are  claimed to show superparamagnetic or spin glass like
 behavior. \cite{Gruyters,Tiwari,Ghosh,Makhlouf,Gang,Zhang,Winkler}
Cupric oxide (CuO) is different from other transition metal monoxides magnetically and its magnetism is perhaps the least understood among them, showing some sort of magnetic order even above its Néel temperature. Because of this it was felt that a magnetic study of the  nanoparticles of CuO may turn out to be very interesting.

  Bulk CuO  has attracted some attention due to its
structural resemblance to high $T_{c}$ superconductors.   It has
been known experimentally by neutron scattering,
specific heat and magnetic susceptibility studies that CuO  undergoes a transition to an
incommensurate antiferromagnetic state  at its Néel temperature
 230 K followed by a transition from the incommensurate to a
commensurate antiferromagnetic state at 213 K. \cite{Yang,Ota,Junod}
However, strangely, the magnetic susceptibility of CuO instead of peaking at its Néel temperature
  undergoes  a change in slope there and shows a broad maximum
  at about 540 K.  \cite{Keeffe} This behavior has been claimed by many authors
  as a manifestation of its quasi one dimensional nature and the related presence of some sort of
short range order above the Néel temperature. \cite{Yang,Junod,Arbuzova} There have been claims
 that CuO can be visualized to have a spin fluid state above the Néel temperature where
 the spins are thought to be dynamically correlated over several lattice spacings. \cite{Muraleedharan} The low temperature susceptibility of CuO shows diverse results and this has
been attributed to the existence of paramagnetic defects
like oxygen vacancies.  \cite{Chandrasekhar}

There have been a few  studies on the magnetism of CuO nanoparticles.  The
temperature dependence of magnetization and susceptibility as
reported by various authors are usually  somewhat different and  sometimes
even contradictory.\cite{Punnoose,Ahmad,Rao,Zheng}  Punnoose et al. have studied exchange bias in CuO nanoparticles of various sizes. They have claimed that  6.6~nm particles  show weak ferromagnetism below 40 K and that for particles above 10 nm in size the behavior is almost bulk-like with a reduction in the Néel temperature.\cite{Punnoose}  Various values of Néel temperatures have been estimated for various particle sizes using susceptibility, exchange bias and muon spin resonance studies. \cite{Punnoose,Ahmad,Zheng,Stewart} Bifurcation in FC and ZFC magnetization and hysteresis have been observed in CuO nanoparticles but no effort has been made to systematically study such non-equilibrium behavior.\cite{Punnoose,Ahmad,Zheng,Stewart,Rao,Borzi} In this work we would like to make a systematic study on the non-equilibrium behavior of CuO nanparticles which would address the following issues: (a) What is the nature of the non-equilibrium state of CuO nanoparticles? (b) Does it show superparamagnetic or spin glass-like behavior? (c) Does it show aging and memory effects?

\section{EXPERIMENTAL DETAILS}

CuO nanoparticles were prepared by the precipitation-pyrolysis
method using starting materials of 99.99\% purity.\cite{Fan} The samples were characterized by X-ray
diffraction (XRD) using a Seifert diffractometer
 with Cu~K$\alpha$ radiation.   The particle sizes as determined by XRD using the
Scherrer formula are 9~nm, 13~nm and
  16~nm for samples heated at 250$^{\circ}$C, 300$^{\circ}$C  and 350$^{\circ}$C  respectively.
  The bulk sample (99.9999\%) was bought from Sigma Aldrich pvt Ltd.
All measurements were performed in a SQUID magnetometer (Quantum Design Model MPMS XL5).
 The field cooled (FC) and zero-field-cooled (ZFC) magnetization measurements were
 done in the temperature range 10~K to 300~K.  The FC measurements were done while
  cooling(FCC) as well as while heating(FCW).  Hysteresis measurements were done
  at 10~K, 100~K and 300~K.   Time dependence of thermoremanent magnetization was
  done at temperatures 10~K, 50~K, 100~K, 150~K, 225~K and 300~K for all the samples.
   Memory experiments were done in a field of 250~Oe in both FC and ZFC protocols.

\section{RESULTS AND DISCUSSION}

\subsection{Temperature dependence of magnetization}
\begin{figure}[!b]
\begin{centering}
\includegraphics[width=1\columnwidth]{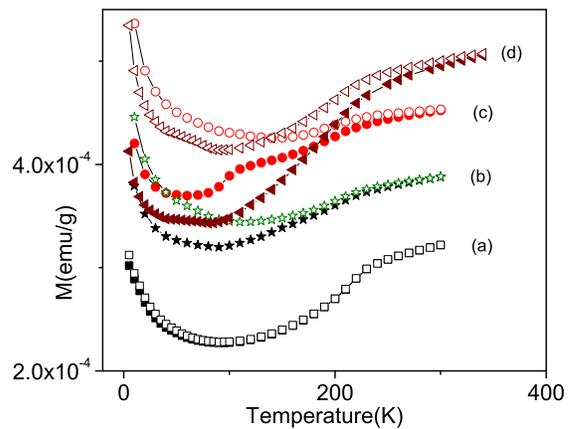}
\par\end{centering}

\caption{(Color online)
FC and ZFC magnetization for (a)bulk, (b)13~nm,(c)9~nm and (d)16~nm samples at 100~Oe.
 Clear bifurcation in FC and ZFC curves can be seen in all the nanoparticle samples.   }
\label{fig:MvsT}
\end{figure}

The temperature dependence of magnetization was done under FC and ZFC protocols
 for all the samples at a field of 100~Oe. See Figure \ref{fig:MvsT}.  It can be seen that
 the FC and ZFC magnetization curves bifurcate above 300~K for the nanoparticle samples,
 while they almost coincide for the bulk sample. No difference between
 FCC and FCW (same as FC) magnetization was seen in any of the samples.
  The magnetization for 16 nm particles is unexpectedly greater than that of 9 nm
  and  13 nm particles.  The reasons for this are not clear, but perhaps this may be due to surface roughness which is not necessarily a monotonic function of particle size.\cite{Sunil}


\begin{figure}[!b]
\begin{centering}
\includegraphics[width=1\columnwidth]{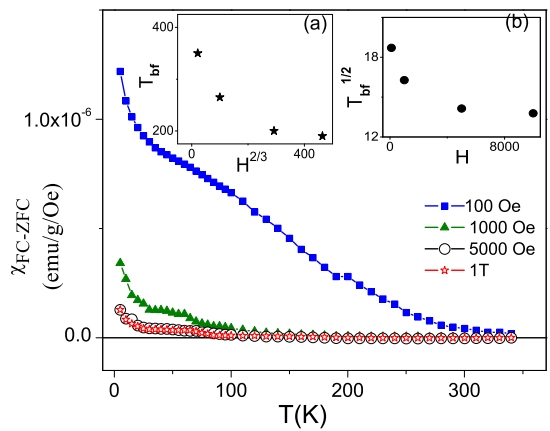}
\par\end{centering}

\caption{(Color online)
 Difference between FC and ZFC susceptibility  for 16~nm nanoparticles for various
  fields: (1) 100~Oe, (2) 1000~Oe, (3) 5000~Oe and (4) 1~T. The insets show the plot of (a)T$_{bf}$  vs $H^{2/3}$ and (b)$T_{bf}^{1/2}$  vs $H$.}

\label{fig:MvsT at various fields}
\end{figure}

Most of the nanoparticles of antiferromagnetic materials have been known to show superparamagnetic  or spin glass-like behavior, well below the Néel temperature of the bulk material, both of which  are characterized by a peak in the
 ZFC magnetization at low fields without a corresponding peak in the FC magnetization. \cite{Ghosh,Tiwari,Salah,Salaha} As can be seen from Figure \ref{fig:MvsT} there is no peak in the ZFC magnetization of CuO nanoparticles. Initially the magnetization decreases with decreasing    temperature and then it increases at low temperatures showing a clear minimum, for all particle sizes.  We would like to check whether this system  shows any   signatures of superparamagnetic or spin glass-like behavior. For this purpose, FC and ZFC magnetization measurements were   done at fields of 100~Oe, 500~Oe, 1000~Oe and 1~T for 16 nm particles. For superparamagnets, the field dependence of the peak temperature, $T_p$, is given by  \cite{Bitoh}
 \begin{equation}
\label{Super-para-blocking}
T_p \propto V\left(1-\frac{H}{H_{K}}\right)^{2},
\end{equation} 
where $V$ is the volume of a particle and $H_K$ is a constant. For spin glasses the corresponding relation is \cite{Almeida}
\begin{equation}
\label{Almeida-Thouless}
H \propto \left(1 - \frac{T_p}{T_f}\right)^{3/2},
\end{equation}
where $T_{f}$ is the spin glass transition temperature in zero applied field. We see that for a superparamagnetic system $T_p^\frac{1}{2}$ is linearly related to $H$ whereas for a spin glass system $T_{p}$ decreases linearly with $H^{2/3}$.

In our case, the CuO nanoparticles do not show any peak in the ZFC curves. Now, in the case of canonical spin glasses, the bifurcation temperature (T$_{bf}$ ) between the FC and ZFC curves and the peak temperature of the ZFC magnetization are very nearly the same.\cite{Mydosh} In superparamagnetic particles it is generally seen that $T_{bf} >  T_p$.\cite{Makhlouf-2}
  $T_{bf}$  can be considered as the onset of superparamagnetic blocking
 or spin glass freezing and so it is expected that $T_{bf}$ will behave in a manner similar to peak
 temperature. This has been shown to be the case for NiO nanoparticles.\cite{Tiwari} Thus, following the example of NiO we shall consider  $T_{bf}$  as a good enough replacement for $T_p$ for further analysis.

        In Figure \ref{fig:MvsT at various fields} we show the difference between FC and ZFC susceptibilites, $\chi_{FC-ZFC}$,  of the 16~nm sample at various applied fields. It is clear that the bifurcation temperature decreases as the applied field increases. For operational reasons,  we shall define the bifurcation temperature as the
 temperature at which $\chi_{FC-ZFC}$ is 1\% of its maximum value.  From  the insets of Figure \ref{fig:MvsT at various fields}, where we look at the functional dependence of $T_{bf}$ on $H$,
  we find that CuO nanoparticles neither follow superparamagnetic nor spin glass behavior.

\subsection{Hysteresis measurements}
We have done hysteresis measurements for all the samples at
temperatures 10~K, 100~K and 300~K. The bulk sample does not show any hysteresis  at
any of the above temperatures but the nanoparticle samples do show hysteresis. We show the coercivity and remanence data for all the nanoparticle samples
in Table \ref{tab:Hysteresis parameters}.  They show
a small hysteresis at temperatures below as well as above the Néel
temperature of the bulk sample.

Hysteresis has been
observed in transition metal monoxide nanoparticles such as  NiO,
MnO, CoO etc. at temperatures below the bulk Néel temperature. \cite{Ghosh,Tiwari}  It has been
attributed to small ferromagnetic contributions due to
uncompensated surface spins. It can be seen that hysteresis is
present even at temperatures above the Néel temperature in CuO
nanoparticles, which is quite unusual, and the origin of this is most likely the short range magnetic order which is present above the Néel temperature.\cite{Yang,Junod,Keeffe}

The fact that this short range order does not lead to any hysteresis in the bulk material means that it is antiferromagnetic in nature. This short range antiferromagnetic order gives rise to a weak ferromagnetism in the nanoparticles, which in turn causes the observed hysteresis even at room temperature, by the mechanism suggested by Neel.\cite{Neel1} This also gives us a lower limit on the scale of the short range order; since this order exists over a region of the size of the nanoparticles the length scale of the short range order should be at least 16~nm, the maximum particle size where we have seen magnetic hysteresis.

\begin{table}[!t]
\begin{centering}
\begin{tabular}{|c|c|c|c|c|c|c|c|c}
\hline
\# & Particle size(nm) & T(K) & Coercivity(Oe) & Remanance(emu/g) \tabularnewline
\hline
\hline
1 & 9 nm & 10 & 38 & 1.  83E-4  \tabularnewline
\hline
2 & 9 nm & 100 & 42  & 1.80E-4  \tabularnewline
\hline
3 & 9 nm & 300 & 55  & 2.30E-4  \tabularnewline
\hline
4 & 13 nm & 10 & 28 &1.  08E-4\tabularnewline
\hline
5 & 13 nm & 100 & 32 &1.  09E-4\tabularnewline
\hline
6 & 13 nm & 300 & 30 & 9.  3E-5\tabularnewline
\hline
7 & 16 nm & 10 & 15 & 6.  85E-5\tabularnewline
\hline
8 & 16 nm & 100 & 60 & 2.26E-4\tabularnewline
\hline
9 & 16 nm & 300 & 24 & 1.  25E-4\tabularnewline
\hline
\end{tabular}
\par\end{centering}

\caption{\label{tab:Hysteresis parameters}Hysteresis parameters for various particle sizes.  }
\end{table}

\subsection{Time dependence of thermoremanant magnetization }
\begin{figure}[!t]
\begin{centering}
\includegraphics[width=1\columnwidth]{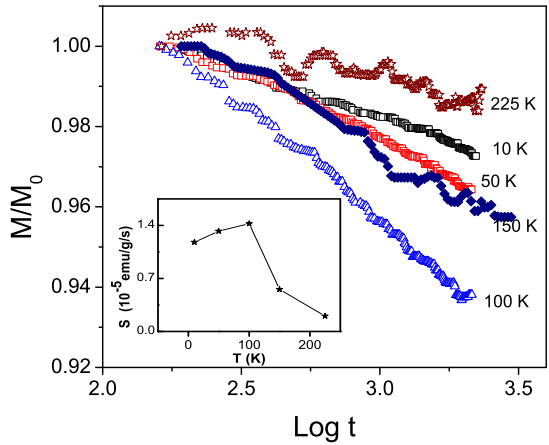}
\par\end{centering}

\caption{(Color online)
Time dependence for 9 nm particles at various temperatures.The data have been smoothened by median filtering. }

\label{fig: Time dep 9 nm}
\end{figure}

Time dependence of thermoremanant magnetization has been measured in all the
samples at various temperatures (10~K, 50~K, 100~K, 150~K, 225~K, and 300~K).
For this measurement a magnetic field of 1.0~kOe was applied
and the sample was cooled to the temperature of interest.   The magnetic
field was now reduced to zero and the magnetization was measured as a function of time.
 We find no time dependence for the bulk sample, but all the nanoparticle samples show
 time dependence at all the temperatures at which the measurements were done.
   In Figure \ref{fig: Time dep 9 nm} we present the time dependence data for the  9 nm particles and it
   is seen to be more or less a logarithmic decay. In the inset of Figure \ref{fig: Time dep 9 nm}, the magnetic
   viscosity (d$M$/d$lnt$) is plotted, which shows a maximum at
   100~K. Such behavior of magnetic viscosity has been observed in other nanoparticle systems as well.\cite{Tiwari,LuO,Guy,Tejada}


\begin{figure}[!t]
\begin{centering}
\includegraphics[width=1\columnwidth]{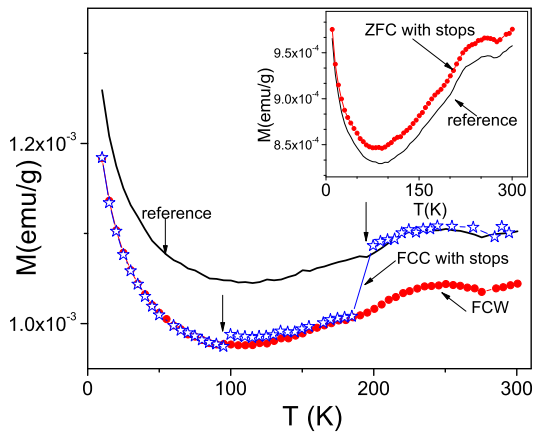}
\par\end{centering}

\caption{(Color online)
Memory experiments done during FC for 16~nm nanoparticles with stops at 100~K and 200~K.  Inset shows the curve for memory experiments during ZFC.   Solid lines are reference curves measured while heating  without any stops in the corresponding cooling process.  }

\label{fig: memory}
\end{figure}

\subsection{Memory Experiments}
Both superparamagnets and spin glasses are known to show memory
effects.\cite{Sasaki,Sun} Superparamagnets are expected to show
FC memory while spin glasses are expected to show memory in both ZFC and FC
protocols.\cite{Sasaki}  Memory experiments were done on the 16 nm sample according to the following protocol.
 Apply 3~T field to the sample for 5
minutes at 300~K to, hopefully, wipe out all memory to begin with. For ZFC
memory,  cool the sample in zero field to 10~K with intermittent
 stops of one hour at 100~K and 200~K.  Measure the magnetization in a field of 250~Oe while warming up to 300~K.  For FC memory,
  cool the sample in an applied field  of 250~Oe with intermittent stops of one hour duration
   at 100~K and 200~K, with the field switched off.   Cool to 10~K finally and then measure the magnetization
   in an applied field of 250~Oe as the temperature is ramped up to 300~K.
    In Figure \ref{fig: memory} we show the results of our memory experiments. It is clear that  no memory effects
    are present in  either
    FC or ZFC protocols. This behavior is in sharp contrast to superparamagnetic or spin glass
   systems which are expected to show memory.



\section{Conclusion}
The magnetic properties of CuO nanoparticles are entirely
different from other antiferromagnetic nanoparticles, for
instance, the usual peak present in ZFC magnetization is absent.
However there is a bifurcation between FC and ZFC magnetization
curves which starts, surprisingly, well above the Néel
temperature. We have shown that this bifurcation does not have
anything to do with the usual spin glass-like or superparamagnetic
behavior shown by other such nanoparticles.
 No memory has been observed in either FC or ZFC protocols which again leads to the very
 strange conclusion that this system behaves neither as a superparamagnet nor as a spin glass.
We believe that the bizarre behavior of CuO nanoparticles originates from the short range
 magnetic order present in the system above the Néel temperature. The short range order
 probably overwhelms the usual manifestations of nanoparticle magnetism.
  Still, the observation of relaxation of magnetization and the associated
   peak in magnetic viscosity is similar to such behavior shown by other nanoparticle systems.

\begin{acknowledgments}
VKB thanks the University Grants Commission of India for financial support.
\end{acknowledgments}

\end{document}